# Continuous focal translation enhances rate of point-scan volumetric microscopy

**COURTNEY JOHNSON,**[1] **JACK EXELL,**[1] **JONATHON KUO,**[1] **AND KEVIN WELSHER**[1*]

[1]*Department of Chemistry, Duke University, 124 Science Dr., Durham, NC 27708, USA*
*\*kevin.welsher@duke.edu*

**Abstract:** Two-Photon Laser-Scanning Microscopy is a powerful tool for exploring biological structure and function because of its ability to optically section through a sample with a tight focus. While it is possible to obtain 3D image stacks by moving a stage, this per-frame imaging process is time consuming. Here, we present a method for an easy-to-implement and inexpensive modification of an existing two-photon microscope to rapidly image in 3D using an electrically tunable lens to create a tessellating scan pattern which repeats with the volume rate. Using appropriate interpolating algorithms, the volumetric imaging rate can be increased by a factor up to four-fold. This capability provides the expansion of the two-photon microscope into the third dimension for faster volumetric imaging capable of visualizing dynamics on timescales not achievable by traditional stage-stack methods.

## 1. Introduction

Confocal laser-scanning microscopy (CLSM) [1, 2]-[3] and two-photon laser scanning microscopy (2P-LSM) [4, 5] have revolutionized live cell imaging and are now the workhorses of biological microscopy. The optical sectioning of both CLSM and 2P-LSM enable 3D microscopy, however, extension into the third spatial dimension results in slower acquisition speeds due to the multiplicative scaling of the number of pixels being acquired. The acquisition of 3D images is conventionally performed using a motorized stage to reposition the sample or objective lens to different focal depths within the sample (hereafter referred to as "stage-stack"). Raster-scanned 2D images are then acquired in a serial fashion to construct a 3D volume (Fig. 1(a,b)). The temporal limitations of this method are that volumetric imaging speed is ultimately driven by two factors: the number of image planes contained within the volume and the speed at which those frames can be acquired, which is ultimately limited by the rate of pixel acquisition.

A further complication of stage-stack 3D imaging is that the time between acquisition of adjacent pixels differs by logarithmic timescales across the three axes in the volume. While the time between scanning two adjacent pixels within a line occurs at the microsecond scale, the time between lines is on the millisecond scale, and the time between adjacent frames is typically on the order of 1 second. Voxels farther apart spatially are thus sampled across even longer temporal intervals. For example, voxels located at the same (*x,y*) coordinate in the top- and bottom-most image planes in a volume are separated by the number of frame-times between them, a time period of 10s of seconds or longer depending on volume size. These challenges make it difficult to image dynamics in 3D using point-scan microscopy. Overcoming these barriers requires new approaches to these tried-and-true stage-stack scanning methods.

Alternative scan patterns have been explored to enhance the speed of 2D scanning microscopy methods through sub-sampling, or acquiring fewer pixels per image. One approach, Random Access Scanning, uses *a priori* knowledge of the specimen location in the image plane to drive a series of galvanic mirrors in an (*x,y*) pattern to image only the pixels the sample is believed to occupy. While this method avoids sampling areas not occupied by a sample it does not provide speed increases for samples where the entire field of view is occupied or where *a priori* knowledge of the sample is unavailable [6-8].

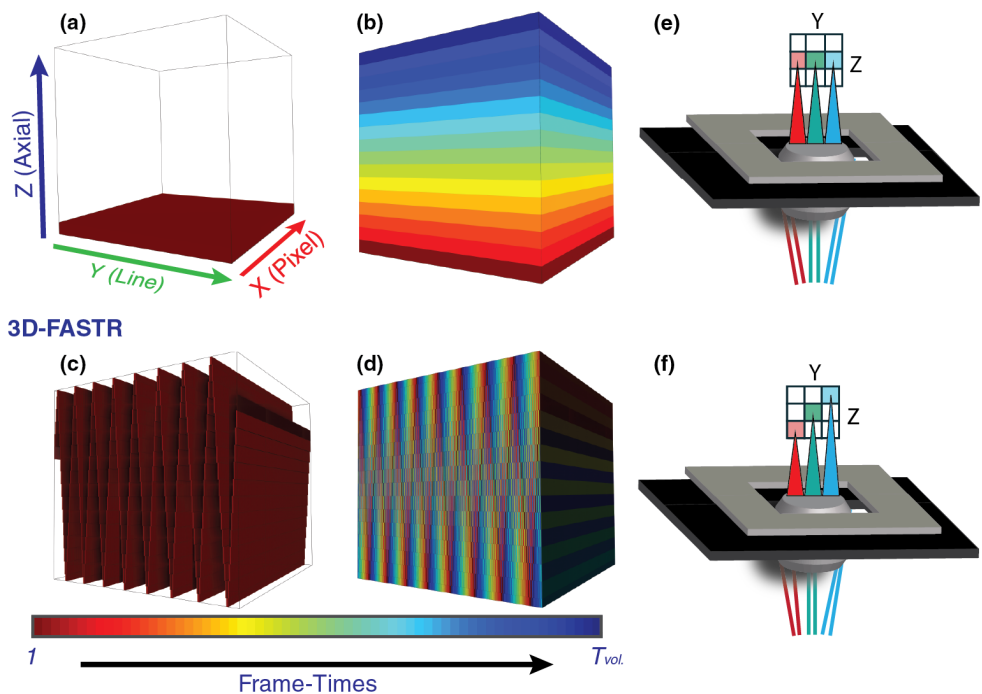

Fig. 1. Comparison between volume scan patterns over time using conventional stage step method versus 3D-FASTR. These example volumes are size 512x512x16 and are shown stretched in Z to a cubic aspect ratio for ease of viewing. (a,c) 3D-view of difference in scan patterns after the first frame-time. While only the lowest image plane has been completely scanned in a conventional stage stack, a triangle wave pattern is apparent with 3D-FASTR, demonstrating sampling across all three axes. (b,d) 3D-view of difference in scan result after the final frame-time, which shows the volume is completely scanned after the passing of 16 frame-times, but the temporal distribution of the scan varies between the two and is color-coded by what frame-time the corresponding voxels were sampled at. (e,f) Cartoon representation of microscope demonstrates YZ scan differences between conventional volumetric acquisition and 3D-FASTR.

Another sparse-sampling scan pattern uses two galvanic mirrors driven sinusoidally to scan the image in a Lissajous pattern. In these methods, the two-dimensional pattern is generated by the synchronization of the two scans such that the pattern is repeated every frame, resulting in the same pixels being sampled repeatedly in time while the unsampled pixels are interpolated. This approach has been demonstrated for Stimulated Raman and Atomic Force Microscopy, both of which are scanning methods [9-11]. A major drawback of this method is that repetition of the Lissajous pattern leads to oversampling of the same regions, particularly at the edges of the frame where the sine wave samples more frequently.

In this work, we improve upon this idea to generate a reproducible 3D pattern which fully and efficiently scans a volume in the fastest theoretically possible time without repeating until the volume is complete. The pattern is generated based on the optimized interaction between three linear waveforms, specifically the interaction between a 2D raster scan with a linear focal displacement. This method, called 3D Fast Acquisition Scan by z-Translating Raster (3D-FASTR), is then implemented through modification of a commercial confocal microscope with the addition of an ultrafast laser system for two-photon excitation and an electrically tunable lens (ETL) to create an easy-to-implement fast 3D imaging system, capable of improving volumetric imaging rates up to four-fold with appropriate interpolation algorithms.

## 2. Theoretical basis for multi-dimensional scanning

### 2.1 Linear axial scanning

It is convenient to think of a 3D image stack as a series of 2D frames spatially separated along the optical axis (Fig. 1(a,b)). Note that we describe frames in terms of time, *T*, and planes in terms of individual 2D frames located at different positions along the Z axis. Neglecting the time required to move a stage between planes, the amount of time required to completely scan an image stack ($T_{vol}$) is given by Eq. (1). This volume acquisition time depends on the number of planes in the image stack ($N_z$) and the two-dimensional frame-rate ($F_{xy}$). The frame-rate is determined by the total number of pixels per frame ($N_{xy}$) and the average pixel dwell time, *P*, as shown in Eq. (2).

$$T_{vol} = \frac{N_z}{F_{xy}} \tag{1}$$

$$F_{xy} = \frac{1}{N_{xy} \times P} \tag{2}$$

In a conventional CLSM or 2P-LSM volume, 2D frames are acquired sequentially, meaning that the top and the bottom of the volume are temporally separated by nearly $T_{vol}$ (Fig. 1(b)). This temporal discrepancy can be overcome by introducing a continuous linear axial translation, (LAT) during the frame scan, resulting in a 3D pattern which samples each plane during a single 2D frame-time. Fig. 1(c) shows a simulated volume featuring LAT after the first frame-time. When compared to the first 2D frame-time in the conventional stage stack, it is obvious that the scanned voxels are more evenly distributed across the depth of the volume. This results in a more even sampling of the volume over time (Fig. 1(d)). While the stage-stack scans a unique plane each frame-time, in the LAT case all image planes are scanned in a different (*x,y*) location every frame-time, drastically reducing the sampling time between different planes within the volume. This idea is illustrated further in the YZ cartoon representation shown in Fig. 1(e) which demonstrates how a conventional stage stack scans each line in a frame at the same focal plane before moving on, while the 3D-FASTR microscope illustrated in Fig. 1(f) changes focal planes at different lines within the same frame scan.

### 2.2 Optimal volume filling conditions

When introducing a linear axial scan during a traditional 2D raster, it is critical to optimize the relative scan rates to ensure unique and even sampling of the volume. When the ratio of the 2D frame and LAT scan frequencies, *R*, meet certain conditions, each voxel will be sampled once and only once. The volume sampling behavior is described by Eq. (3):

$$R = \frac{F_z}{F_{xy}} = n + \frac{m}{N_z} \tag{3}$$

Here $F_z$ is the frequency of the LAT and is greater than $F_{xy}$, the frequency of the 2D frame scan (frame-rate). $N_z$ is the number of sections along the z-axis. The value *n*, which we refer to here as the fundamental, is the base integer ratio of the axial scan to the 2D scan. The value of $n$ is derived from the Euclidean (integer) quotient of $F_z$ and $F_{xy}$. The value $m$, which we denote as the shift parameter, represents the amount of phase shift in the Z-scan between frame-times. Repeatable scan patterns can be achieved for any integer value of $m$ in the ratio $m/N_z$. Volume fill efficiency is completely dictated by Eq. (3), illustrated here in Fig. 2(a), which shows how the quantity of unsampled voxels remaining after the volume should theoretically be scanned to completion ($T_{vol}$) depends on the ratio of the axial and 2D scan frequencies. To achieve this minimum completion time, all acquired pixels must be uniquely located in the volume space so that no voxel positions are multiply sampled; otherwise the actual volume completion time will be longer. It is clear from this plot that the experimental parameters must be very carefully chosen to minimize the number of unsampled voxels at $T_{vol}$. Consider a few illustrative examples.

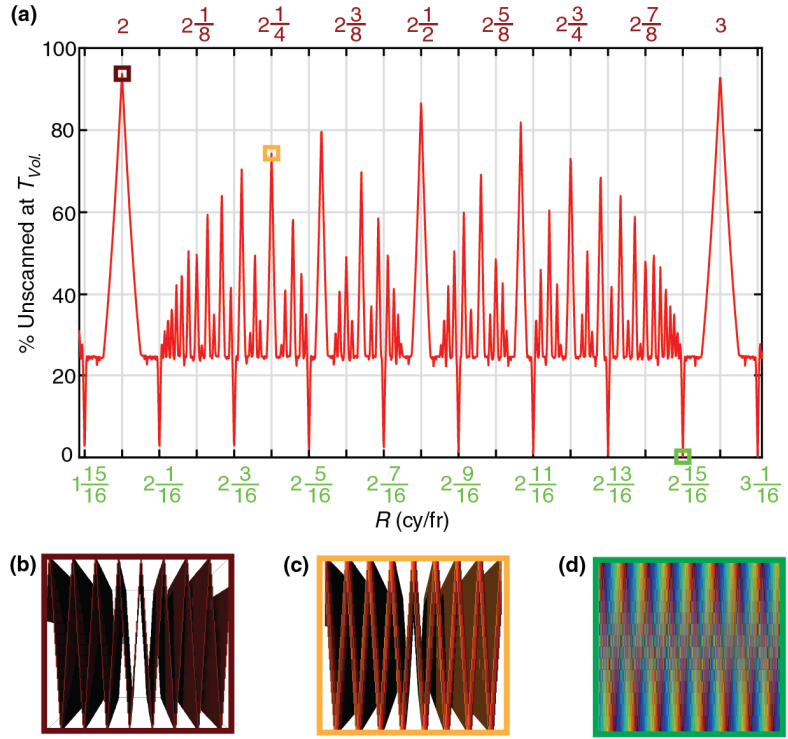

Fig. 2. (a) Theoretical relationship between number of Z-translation cycles per frame-time ($R$), and fill efficiency, demonstrating the effect of pattern timing on volume completion after $T_{vol}$ for a simulated volume consisting of $N_z = 16$ image planes. The effects of $R$ on volume completion are illustrated using the YZ cross-section of the color-coded scan map as seen in Fig. 1(c-d). Here, 3 examples designated by the boxed peaks correspond to (b). global maxima (red), (c). local maxima (yellow), and (d). global minima (green).

### 2.2.1 m = 0

In this case, the LAT frequency is an exact multiple of the 2D frame-time such that the ratio, $R$ is an integer equal to the fundamental, $n$. This condition represents synchronization between the two scans because there is zero relative phase shift over time, meaning each 2D frame begins at the same point within the LAT scan. This corresponds to the worst possible volumetric fill rate as the number of unique scanned voxels will not progress further than a factor of $1/N_z$. Take as an example a scan where $F_z = 4$ Hz and $F_{xy} = 1$ Hz, which yields an $R = 4$, with fundamental of $n = 4$ and a shift of $m = 0$, meaning the pattern repeats each 2D frame-time, regardless of the number of z-planes. These values correspond to the global maxima in terms of the percentage of the volume which is unsampled. These maxima can be clearly seen in Fig. 2(a), with an example volumetric cross-section of this repeating pattern shown in Fig. 2(b).

### 2.2.2 $m/N_z$ is a reducible fraction

The ratio of the shift variable, $m$, and the number of z-sections plays a critical role in determining the efficiency with which the volume is sampled. When $m$ is an integer, the scan pattern will phase shift $m$ z-sections each 2D frame-time. For example, if a volume has 16 sections and $m = 2$, then the scan will shift two z-sections each 2D frame-time.

The lowest common denominator of the ratio of the shift number, $m$, to the number of z-sections, $N_z$, indicates the number of 2D frame-times before the pattern repeats itself and samples the same voxel a second time. For example, if $F_z = 4.125$ Hz and $F_{xy} = 1$ Hz for a volume with $N_z = 16$, Eq. 3 is satisfied as $R = 4 + 2/16$ with $n = 4$ and a remainder of 2/16. This yields a shift value of $m = 2$, meaning each 2D frame-time the z-section of a particular $(x,y)$ pixel location will shift by two z-bins. Reducing 2/16 yields 1/8, meaning that the entire pattern will repeat after 8 frames. The number of voxels sampled will peak after 8 elapsed frame-times and will never sample new voxels, instead scanning the same voxels a second time over during the next eight 2D frame-times. While this condition samples the volume better than for shift number $m = 0$, it is still far from an ideal sampling. These points show up as local maxima in Fig. 2(a), with an example cross-section shown in Fig. 2(c) that demonstrates the highest possible fill when the remainder $m/N_z$ equals 1/4.

### 2.2.3 $m$ and $N_z$ are coprime

The volumetric filling efficiency is optimized when $m/N_z$ is an irreducible fraction. Following on from the example above, consider $F_z = 4.0625$ Hz and $F_{xy} = 1$ Hz for a volume with $N_z = 16$. From Eq. 3, this yields $R = 4 + 1/16$ with a fundamental of $n = 4$ and a shift number $m = 1$, meaning each 2D frame-time the scan shifts by one z-bin. The denominator of the remaining fraction in its reduced form is 16, meaning each voxel will be sampled once every 16 2D frame-times, just as it would be in a traditional stage stack. This condition corresponds to the minima in Fig. 2(a), where all voxels are sampled at a time equal to $T_{vol}$. A cross-section of an example filled volume can be seen in Fig. 2(b).

## 3. Implementation of 3D-FASTR

The model presented here requires only a point-scan microscope and a method of Z-translation. The LAT could be accomplished in many ways either optical or mechanical. The specific implementation described here uses an electrically-tunable lens (ETL) which is an inexpensive and mechanically non-perturbative option for achieving dynamic focal changes. An ETL consists of a fluid-filled elastic membrane which deforms as a function of applied current. These varifocal lenses have been utilized to add volumetric imaging capabilities to existing microscopes such as light-sheet [12], particle tracking [13], temporal focusing [14], miniaturized two-photon endoscopes [15], or to generate structured illumination microscopy patterns [16, 17]. The ETL model featured here (ETL 10-30-C, Optotune) is capable of deforming from a minimum focal power of -2.2 dpt to a maximum of +4.5 dpt. Additionally, it is capable of Z-translation frequencies up to 1000 Hz.

### 3.1 Microscope setup

The general layout of the microscope is shown in Fig. 3. The two-photon excitation source was a tunable-wavelength pulsed laser (100 fs, 80 MHz, Chameleon Discovery, Coherent) featuring pre-compensation for group velocity dispersion set to 10,800 fs$^2$ to maximize two-photon fluorescence intensity at the sample [18]. The beam is then steered into a 10:90 beamsplitter (BS025, Thorlabs) with 10% of the laser power focused through a 75mm lens (AC254-075-B-ML, Thorlabs) onto a Silicon photodiode (DET10A, Thorlabs) terminated with a 56kΩ resistor.

The remaining 90% beam passes through the ETL (EL-10-30-C-NIR-LD-MV, Optotune), which is driven by a clean current source (Arroyo Instruments 4200-DR LaserSource) and modulated by a function generator (FG, Stanford Research System DS345). The ETL is positioned approximately 1200 mm from the objective lens and mounted vertically, with the optical axis perpendicular to the table surface to prevent gravity-induced coma as suggested by the manufacturer. The emerging beam is split again by a 90:10 beamsplitter (BS029, Thorlabs).

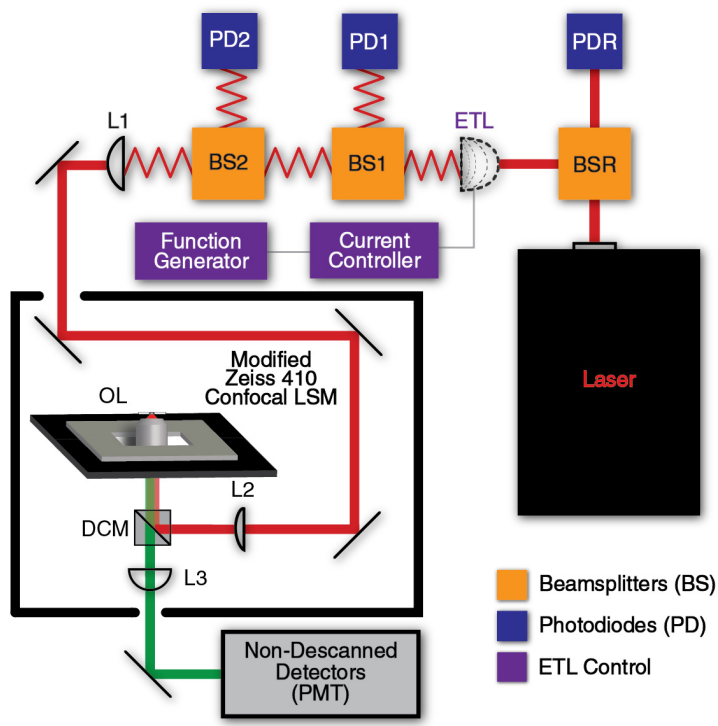

Fig. 3. Instrument diagram for 3D-FASTR implementation. Beam is split using beamsplitters (BS) and measured by photodiodes (PD). Laser power is analyzed as reference by splitting at BSR and measuring at photodiode (PDR) before focal deflection by electrically-tunable lens (ETL). Focus is relayed over the long distance using two lenses (L1/L2) before entering the confocal scanner. The focal range is shifted using L3 before deflection by dichroic mirror (DCM) to objective lens (OL). Emission passes back through DCM to non-descanned detection PMTs.

The 10% beam is directed to a second photodiode (DETA10A, Thorlabs, 56 kΩ termination), intended to maximize response at high ETL focal powers. The remaining 90% continues to another 90:10 beamsplitter. The 10% beam is focused by a 200 mm lens (AC254-200-B-ML, Thorlabs) before occlusion by a 100 μm pinhole positioned 250 mm from the lens in front of a third photodiode (DET100A, Thorlabs, 56 kΩ termination), intended to measure low ETL focal powers.

The 90% beam continues to a 500 mm relay lens (AC254-500-A-ML, Thorlabs) positioned 950 mm from the ETL such that the emerging beam is collimated when the ETL is driven to a focal power +2.5 diopters. The emerging beam then passes through the rear entrance of the modified confocal LSM (Zeiss LSM 410, modified by LSMTech) where it is reflected by an internal 730 nm shortpass dichroic mirror into the scan unit, where the beam is deflected in XY by the scanning mirrors before passing through a slider-mounted 1000 mm offset lens and reflecting off a 700 nm shortpass dichroic mirror up to the objective lens. The emission passes back through the previous dichroic and is focused by an 85 mm tube lens located in the detection side of the slider. The excitation is filtered by a multiphoton blocker (FF01-750/SP-25, Semrock) placed in the detection pathway prior to reaching the non-descanned detection PMT unit which contains a 570nm longpass dichroic, reflecting to a green, 540nm/45nm bandpass filter and passing to a red bandpass filter (FF01-731/137-25, Semrock).

*3.2 ETL focal length detection*

As the volume is not scanned sequentially, the ($x,y,z$) coordinates for each acquired pixel must be determined. While the ($x,y$) coordinates are recorded from the raster scan pixel and line positions, the sampled focal plane depends on the ETL focal power and varies over time. Further, temperature and waveform frequency both affect the output focal power of the ETL for a given current input, making it necessary to have a detection system that reads out the actual focal power in real-time. A system of three photodiodes was used to measure the ETL focal length to fulfill this requirement. The first photodiode is positioned prior to the ETL and acts as a laser power reference (Fig. 3, PDR). This photodiode measures the laser power to correct the reading of the two subsequent photodiodes such that they represent ETL focal power only. The second photodiode (Fig. 3, PD1) has a small detection radius (0.1 mm$^2$) and was positioned approximately 200 mm beyond the ETL, corresponding to the maximum focal power of the ETL. The beam dilation which occurs with increases of the ETL focal length is then measured as a decrease in signal on PD1. The diagram in Fig. 4(a) demonstrates the working principle of this detection method. At long focal lengths, the spot size of the beam at PD1 is much greater than the active detector area, and consequently the sensitivity of this signal change is reduced. This focal power regime also corresponds to the highest rate of change in focal shift at the image plane with respect to changes in current, meaning this control current region requires the highest sensitivity to changes for accurate measurement of focus.

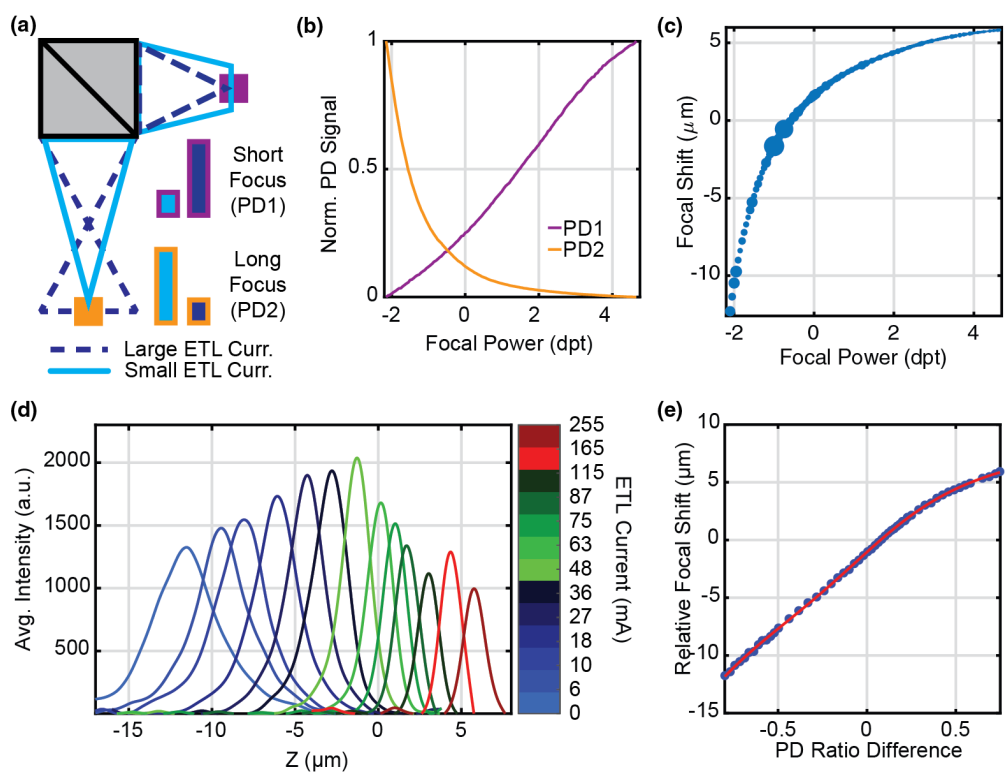

Fig. 4. Detection of real-time ETL focal depth. (a) Principle of photodiode detection scheme shows how two detectors positioned at different distances measure inverse signal levels as a function of ETL focal length. (b) Signal levels for each measurement photodiode as a function of ETL focal power. (c) Relationship between ETL focal power and focal shift in the image plane represented with blue dots scaled in height to match measurement uncertainty. (d) Gaussian-like intensity peak of image stacks at different ETL drive currents (color-coded) show shift in focal depth relative to reference image plane. (e) The signal difference between the two photodiodes shown in blue forms an almost linear relationship as a function of ETL focal power with the final calibration fitting shown in red.

Addition of a third photodiode (Fig. 3, PD2) enables sensitive detection of focal changes at the lowest focal powers. Due to its larger detection area, a 100 μm pinhole is positioned such that the PD2 signal is maximized when the ETL is driven to its minimum focal power of approximately -2.2 dpt. This is accomplished using a 200 mm lens and positioning the pinhole approximately 250 mm from that lens. The normalized response across the ETL focal power range is shown in Fig. 4(b) which demonstrates the inverse response character of the two photodiodes. The higher responsivity of PD3 causes a steeper loss of signal than PD2, yielding greater sensitivity in the drive current regime that produces the largest change in focal depth.

### 3.3 ETL calibration

A calibration must be performed to correlate the signal of the three photodiodes to a real focal shift in the image plane given the nonlinear relationship between the input current and output focal shift (Fig. 4(c)). Calibration data is obtained by collecting volumetric stage stacks of coverslip-bound fluorescent microspheres (Bangs Labs FCSG003, carboxyl-functionalized polystyrene, surf green, 200 nm diameter, 6.7 ng/μL, PBS) at different ETL focal powers. The mean intensity versus stage position for each applied current is fit to a cubic interpolant, where the peak location compared to a collimated reference beam corresponds to the relative focal shift in the image plane (Fig. 4(d)). The FWHM of these curves at different axial positions shows the relative axial resolution as a function of ETL focal power. The resulting plot of focal shift versus mean photodiode ratio difference ($\frac{PD2-PD1}{PDR}$) across all currents yields the calibration curve shown in Fig. 4(e) which can be used to measure the focal depth in real time.

### 3.4 Triangle waveform generation procedure

The model presented here relies on the continuous focal deflection as a linear function to scan with the efficiency expected. As discussed previously, the ETL does not produce focal shifts in the image plane that are linear with focal power. Consequently, a triangular drive signal fails to produce linear focal shifts over time. To correct this, a function generator was used to create a custom input function which yields the desired periodic focal shifts to match the theoretical behavior described above.

### 3.4.1 Waveform creation

Initial data were collected by using the FG to create a triangle current (TC) wave to drive the ETL at the desired frequency and amplitude. FG voltage and photodiode readings were acquired over at least 20 ETL cycles. This base data is used to measure the relationship between input FG voltage and resultant focal shift. The FG voltage versus focal position data were fit with a polynomial interpolant to generate a look-up table (LUT) which maps the FG voltages onto the resulting focal positions (Fig. 5(a)). The LUT is then used to generate an arbitrary waveform (AWF) which yields a triangle wave in focal position.

This waveform correction significantly improves the ability for this 3D-FASTR implementation to achieve the results theorized by the model. Fig. 5(b) demonstrates the input periodic current obtained by the waveform generation procedure outlined above, compared to the default current triangle wave. The generated input wave visibly spends less time in current regions with minimal changes in focus, and more time in current regimes where small changes in current lead to bigger changes in focus. This change compared to the TC input yields a dramatic difference in the resulting output waveforms shown in Fig. 5(c). Here, the generated AWF precisely creates a triangular periodic focal shift, which overlays with the desired theoretical waveform.

This result can be alternatively visualized in the form of the bar chart in Fig. 5(d). Here, a flat-shaped bar chart of axial phases, illustrated by the dotted red line, represents the distribution corresponding to the ideal linear translation model where each focal plane is sampled equally. The difference in plane sampling distributions between an uncorrected triangular drive signal (purple) and an AWF at $R = 3 - (1/16)$ cy./fr. (green) show the impact of correction. The TC is obviously biased toward lower focal planes, but the AWF approaches the ideal model.

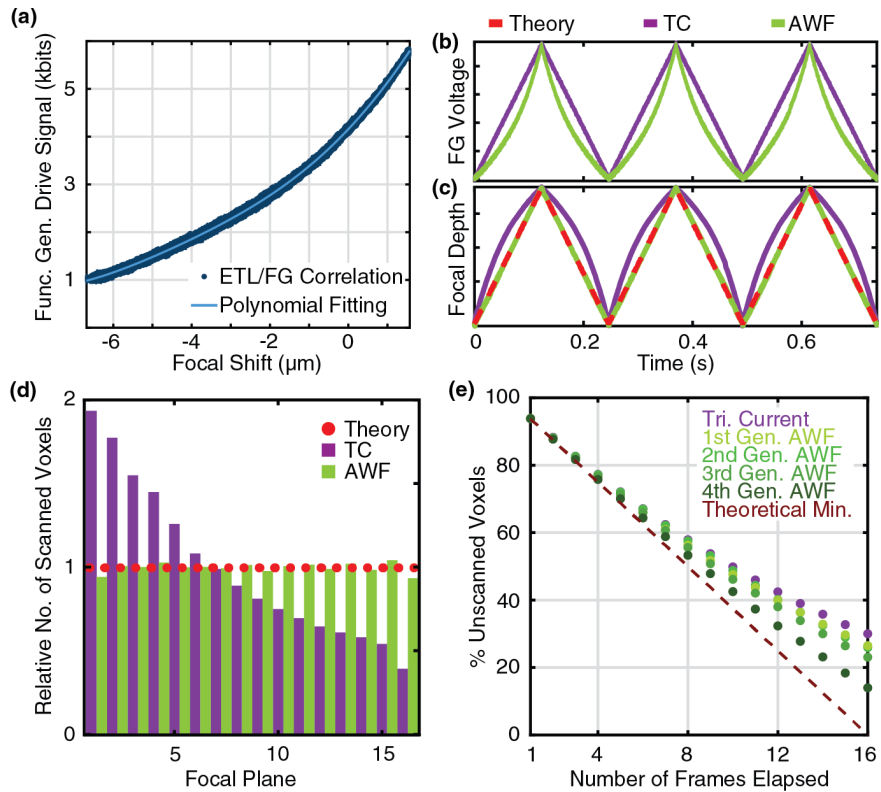

Fig. 5. Impact of arbitrary waveform (AWF) creation on efficiency. (a) Polynomial-fit relationship between input/output of input triangle current (TC) waveform (b). Comparison of ETL drive signal over time between TC and AWF. (c). Comparison of resultant focal shift from TC and AWF. The AWF current pattern at ~4Hz produces a triangular focal shift which closely matches the model, providing optimum fill efficiency. (d). Bar chart of sampled focal planes across an arbitrary time period evaluates linearity of Z-translation by comparing the total number of voxels sampled in each focal plane. A relative value of 1 corresponds to the theoretical model where each focal plane is sampled equally, illustrated as a dotted red line. The uncorrected TC shows bias to lower image planes, while the AWF's performance approaches the model. (e) Improvement in scan efficiency with successive iterations of waveform generation at ~49Hz ETL frequency.

### 3.4.2 Further improvement of waveform efficiency

At ~4 Hz an arbitrary function generated from a triangular input will produce a nearly ideal focal shift, but at higher frequencies the waveform must be refined to produce optimal linearity. The generated AWF can serve as the new input and this process can be repeated in an iterative fashion, improving volume fill performance. As the ETL frequency increases, more iterations will be required to produce optimal linear waveforms as the ability to correct the ETL degrades. This process of data acquisition and fitting is iterated until improvement in the percent of unscanned voxels is no longer observed.

The AWF performance was evaluated at frequencies up to 49 Hz and Fig. 5(e) demonstrates how successive waveform generations at this high frequency can produce increasingly efficient LATs. Fig. 6(a) shows how the ability to produce an ideal LAT declines as ETL frequency increases, resulting in lower volume fill efficiency compared to the theoretical model, but which still outperforms the uncorrected 595 Hz sine wave shown for comparison. The order of the polynomial used to fit the FG/Focus relationship can greatly impact whether the output AWF will be optimally efficient. At low frequencies where the ETL is well-behaved, higher-order fits more accurately represent the correlation, but at high frequencies where the waveform is more erratic, a series of lower-order fits effectively smooth the noisy correlation and produce ultimately better AWFs. Table 1 lists the tested AWF parameters and optimum fitting regimes for a range of ETL frequencies.

**Table 1. Sequencing of Waveform Fits by Frequency**

| ETL Frequency (Hz) | Target R (cy/fr) | Number of Iterations | Best Achieved Fill (Empty) at 1x Volume Time |
|---|---|---|---|
| 4.066969 Hz | 3 – 1/16 | 2 | 92.56% (7.44%) |
| 10.989470 Hz | 8 – 1/16 | 2 | 91.22% (8.78%) |
| 26.218969 Hz | 19 – 1/16 | 3 | 89.81% (10.19%) |
| 48.371054 Hz | 35 – 1/16 | 4 | 86.78% (13.22%) |

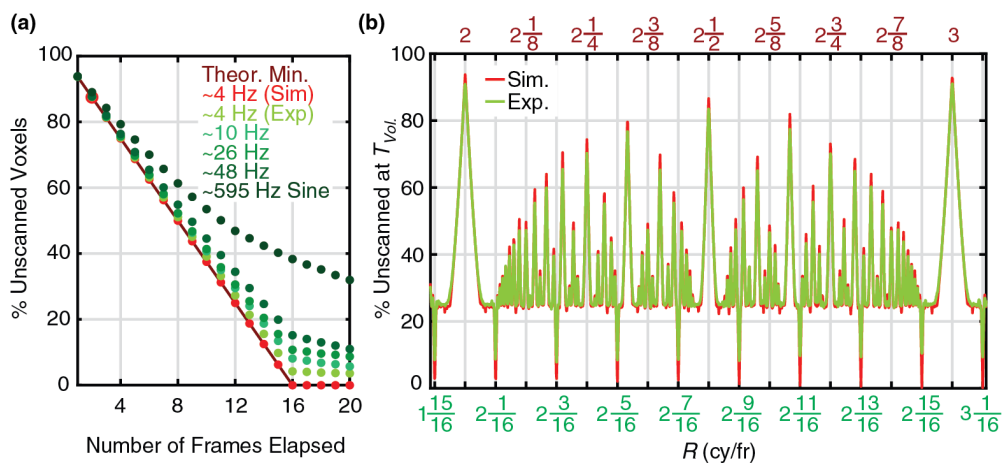

Fig. 6. 3D-FASTR Implementation vs. Theory. (a) Fill efficiency declines with increasing ETL frequency due to decreasing ability to correct ETL waveform, but the result still shows significant improvement compared to an uncorrected sine wave. (b). Low-frequency AWF shows expected pattern timing behavior with respect to $R$ and approaches fill levels of theoretical 3D-FASTR model.

Fig. 6(b) shows that at low frequencies, the ability to correct the ETL produces a waveform which closely adheres to the theoretical model LAT and as a result, the experimental system's behavior replicates the model.

## 4. Live cell volumetric imaging with 3D-FASTR

Here we demonstrate that the 3D-FASTR method developed above and its application leads to significant increases in the speed of volumetric imaging in live cells.

### 4.1 Cell culture

HeLa cells were cultured in FluoroBrite DMEM Media (Life Technologies, #A1896701) supplemented with 10 % fetal bovine serum (Millipore Sigma, #F2442), 1X penicillin-streptomycin (Corning, #30-002-CI), and 1X GlutaMAX (Life Technologies, # 35050061). A day before imaging cells were plated in complete DMEM at 1X $10^5$ cells/well in a 8 well μ-slide, glass bottom (Ibidi, #80827). Cells were maintained at 37 °C with 5% $CO_2$.

### 4.2 Sample preparation

Directly prior to imaging, cells were washed three times with Dulbecco's phosphate buffered saline (DPBS) (HyClone, #SH30264.01) and stained with either 2.5 μM SYTO 41 (Life Technologies, #S11352), or 12.5 μM SYTO 61 (Life Technologies, #S11343) in Live Cell Imaging Solution (LCIS) (Life Technologies, #A14291DJ). The staining proceeded for 30 minutes in the 37 °C $CO_2$ incubator. To enable simultaneous imaging of the nucleic acid content and cell membrane, 12.5 μg/mL DiA (Life Technologies, #D3883) in LCIS was added to the cells stained with SYTO 61 and incubated for an additional 30 minutes. Pluronic F-127 (Millipore Sigma, # P2443), was added to 1 mg/mL DiA in DMSO at a ratio of 1:50 and sonicated for 10 minutes before any dilution to help with solubility. For both the single and double labelling experiments the staining solution was removed, cells were again washed successively three times with DPBS before adding fresh LCIS and placing the sample on the microscope stage for imaging.

*4.3 Imaging process*

As the slower scan, the frame-rate is the driver of absolute volumetric imaging speed, so the maximum value of 1.3845 fr/s was utilized and the ETL frequency was adjusted to achieve the desired $R$ value for $N_z = 16$. Pixels were acquired at 250 nm spatial intervals across $(x,y)$. Images are acquired as a stream of pixels. Pixel number, line number, and photodiode readouts are recorded to determine the $(x,y,z)$ coordinates for each acquired pixel. The $z$-coordinate location for each acquired pixel is determined by converting the recorded photodiode ratio difference to the corresponding focal shift using the calibration curve. These real-space axial positions are then assigned to a discrete plane number. The total number of planes ($N_z$) is critical, as it controls the degree of phase shift each frame. The amplitude of the ETL waveform divided by the desired number of discrete planes gives the distance between axial planes. The mean depth of focal shift peak minima and maxima are used to optimally set the axial boundaries of the volume. This means that some acquired pixels will always be excluded due to detection noise. Optimizing detection precision and drive signal purity can minimize this loss of efficiency and lead to higher fill rates.

After the $z$-coordinate is determined, the voxel is assigned a numeric value based on PMT intensity. Separate volumes are created for each color channel which differ only by the filled intensity value. When inefficiency causes voxels to be scanned more than once, the oversampled voxel is assigned a value based on the maximum PMT reading in each channel.

Fig. 7 demonstrates the performance of the 3D-FASTR microscope on a sample of HeLa cells which were imaged using an LAT with $R = 35 - (1/16)$ cy./fr. across a focal shift range of approximately 8 μm. The cells were dyed with red nucleic acid stain Syto61 and green membrane stain DiA. Fig. 7(a-c) demonstrate the 3D-FASTR imaging process by showing images of three focal planes spaced throughout a volume consisting of 16 image planes. These "scan map" images show the combined red and green intensity values for each sampled voxel, while unsampled regions are highlighted in blue.

These unscanned voxels are inpainted using the MATLAB `inpaintn` function developed by Garcia et al. [19], which draws spatial frequency information from neighboring scanned voxels to complete missing regions. After interpolation, the intensity for each channel is thresholded from a floor PMT value of 0 to 95% maximum recorded PMT intensity, then rescaled to 0-255 to create the final 8-bit RGB image. Fig. 7(d-f) demonstrate the resulting final images after inpainting.

Three-dimensional intensity data was ported into Avizo 9.5 through a binary data file with specified volume and voxel sizes. Both the color and alpha values were calculated to be linear with intensity. To improve edge definition, the Edge 3D post-processing effect was enabled with a Gradient Threshold of 0.0001. Global illumination, including ambient occlusion was enabled to ensure the rendering of realistic lighting. Fig. 7(g) shows the direct volume rendering of the final 3D structure using Avizo. As an alternative display of the 3D profile of the final volume, Fig. 7(h) shows an example XZ sample of this volume.

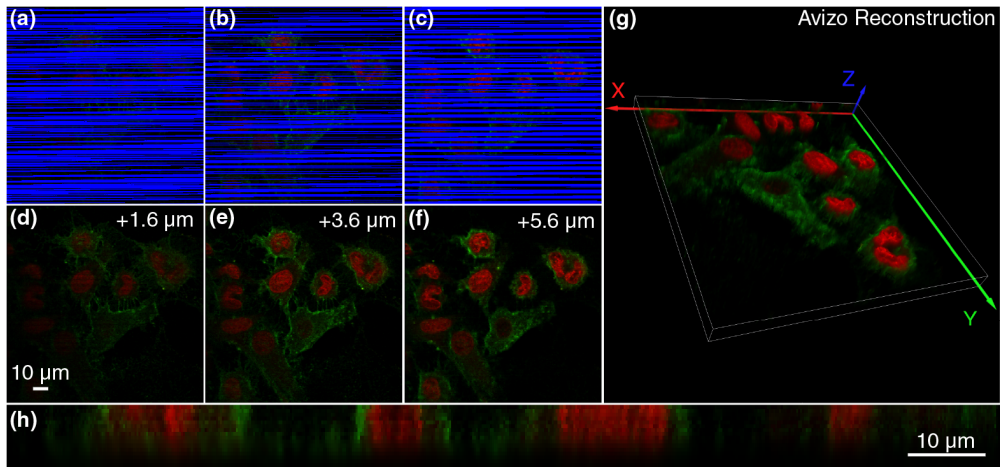

Fig. 7. The imaging process displayed through a series of 3 image planes (7, 11, 15) from a 16 z-plane volume of live HeLa cells stained with red nucleic acid dye Syto61 and green membrane dye DiA. Volume was acquired using a 3D-FASTR pattern with $R = 35 - 1/16$ after 50% $T_{vol}$ (2x speed increase). (a-c) A scan map image showing real intensity of scanned voxels and highlighting unsampled voxels in blue for each image plane. (d-f). Corresponding final interpolated image. (g). Reconstruction of 128x128x8 μm 3D-FASTR volume from 512x512x16 voxels. (h). XZ section shows depth profile of sample across line 187.

*4.4 Image quality factors*

The imaging rate improvement achievable by the 3D-FASTR method will depend on the desired image quality and ability to maximize the coverage of sampled voxels, with image quality tradeoffs required for faster volume acquisition rates. Speed increases arise from the ability to generate a volumetric image in fewer elapsed frame-times, enabled by the sparse sampling behavior of the scan pattern. Thus, the final image quality is highly dependent on the ability of the inpainting algorithm to fill in unsampled regions.

Fig. 8 illustrates this balance between speed and image quality, and how the variables $n$ and $m$ play integral roles in achieving optimal sampling distributions. Fig. 8(a) shows a full field of HeLa cells dyed with nucleic acid stain SYTO 41 located at a focal depth corresponding to the bottom section of a 16 z-plane volume acquired at $T_{Vol}$. The purple outline shows the region of cells which will be used to compare imaging parameters, shown in enlarged form in Fig. 8(b), to serve as a reference.

The quantity of neighboring voxels has a significant effect on the quality of the final image such that the spatial distribution of scanned voxels is more important than the relative quantity filled. This distribution is a function of both the shift number, $m$, and the fundamental, $n$. As discussed previously, the shift number is the amount of phase distance traveled each frame. When the shift number, $m$, is 1, the scanned stripe patterns fill adjacent to each other with no unsampled gaps between scanned regions. This scenario tiles without oversampling voxels, but interpolates poorly because scanned voxels are clustered together in adjacent lines. This effect is exaggerated away from the center planes of the volume, creating large gaps that are unsampled and have no neighboring scanned voxels to draw from for inpainting.

This condition is illustrated in the 3D map in Fig. 8(c), which shows the number of adjacent scanned voxels at each voxel location in a 512x512x16 volume after 50% $T_{vol}$ with $R = 8 + 1/16$ cy./fr. The scan map in Fig. 8(e) shows the uninterpolated image after acquisition for 50% $T_{vol}$. In this condition, the unsampled regions are large, and as Fig. 8(c) reveals, most unsampled voxels also have no scanned neighbors. This lack of neighboring information coupled with large unsampled gaps result in the appearance of blurry bands in the final image.

Increasing the shift number causes greater phase shift over time, creating spatially-offset gaps between scanned voxels in each image plane which achieve more even coverage of scanned voxels by reducing line-length gaps. This is visible in Fig. 8(d), which shows the number of nearest numbers for a shift number of 7 after 50% $T_{vol}$. Unlike in Fig. 8(c), there is no major bias of neighboring voxels at center planes, instead, all planes show relatively comparable distribution of nearest neighbors. This leads to improved image quality as shown in Fig. 8(f). Increasing the shift number results in a greater phase-shift over time which creates offset-gaps between scanned voxels in each image plane. By spreading the scanned voxels over a larger area, this condition achieves a more even distribution and reduces line-length gaps as shown in Fig. 8(f). It is clear here that the large blurry bands seen in Fig. 8(e) are absent, with only a few stray lines with noisy pixels. Increasing the value of $n$ to 35 as shown in Fig. 8(g) removes this noise and provides an optimal quality image at 50% $T_{vol}$.

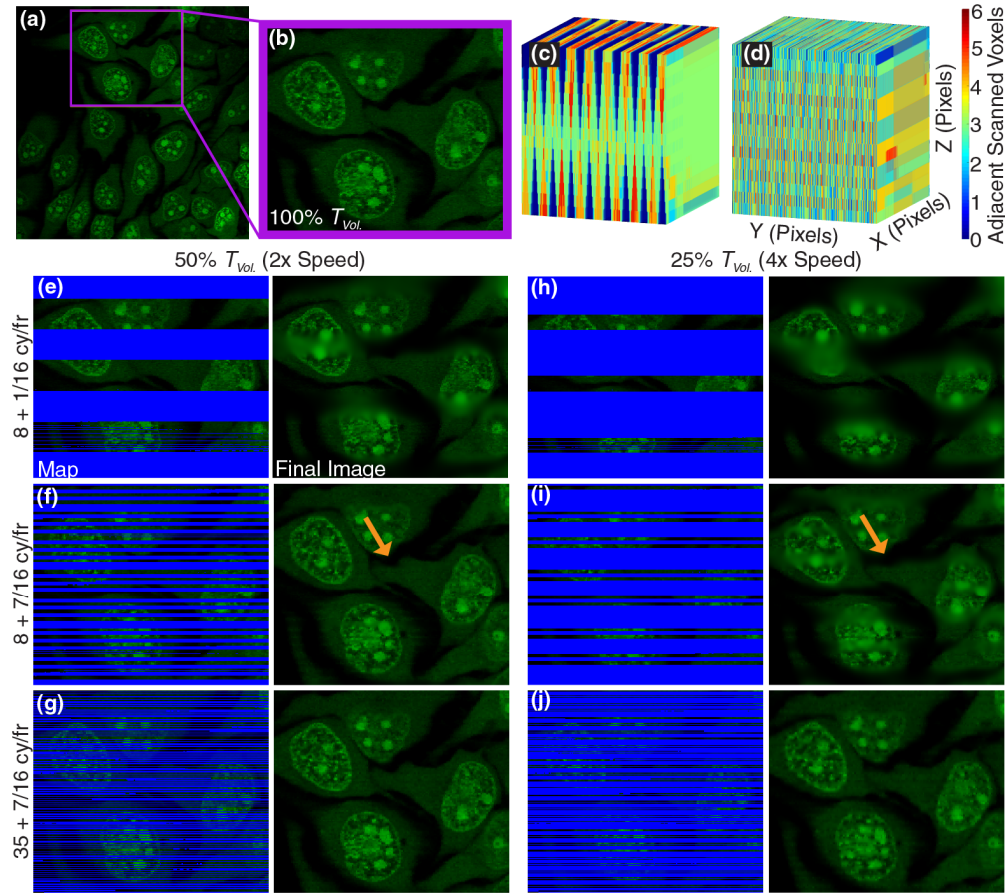

Fig. 8. Trade-offs between imaging speed and fidelity for HeLa cells stained with Syto41 (nucleic acids, green) (a). Full image of bottom plane (depth of -1.25 μm) of volume constructed using 3D-FASTR with crop region outlined in purple. (b). Cropped image at 100% $T_{vol}$ serves as image quality reference. (c/d). Volume representation of neighboring scanned voxels displayed using the MATLAB function `Vol3D` developed by Joe Conti [20]. These volumes show the number of scanned neighbors for each voxel position with a shift number of (c). $m = 1$ vs. (d). $m = 7$. (e-g). Comparison of scanned voxels and final image quality of images acquired at 50% $T_{vol}$ for different values of $n$ and $m$. (h-j). Comparison of scanned voxels and final image quality of images acquired in 25% $T_{vol}$ for different values of $n$ and $m$. Orange arrows in (f) and (i) highlight curvature artifacts caused by inadequate sampling at 25% $T_{vol}$ that is remedied by increasing $n$ from 8 to 35.

Reducing the acquisition time to 25% $T_{vol}$ (4x faster than conventional) decreases the total number of scanned voxels in the volume, which increases the number of voxels which must be interpolated. In Fig. 8(i), this reduction of information causes striping artifacts and morphological distortions. As highlighted by orange arrows, in Fig. 8(f), a curved shape in the cytoplasm is visible in the central region of the image, but in Fig. 8(i), the lack of information causes this curvature to appear straight. While the higher shift distributes voxels across lines, the coverage with fewer voxels is insufficient as the low fundamental $n$ of 8 means that voxels are scanned consecutively at each plane for several lines. Increasing the value of $n$ decreases the focal dwell time, increasing sampling frequency of planes. This can be visualized in the scan map images as a shorter stripe length, or a decrease in the number of consecutively sampled voxels in a plane. As demonstrated in Fig. 8(j), increasing $n$ to 35 + 7/16 from 8 + 7/16 distributes the voxels efficiently enough to remove the distortions and artifacts seen at 25% $T_{vol}$. While the resulting image quality is less sharp than at 50% $T_{vol}$, all coarse cellular features are preserved. Potentially, higher *n* values would enable even faster acquisition rates than those achieved here, which are ultimately limited by the ability to correct the ETL waveform input.

### 4.5 Capturing dynamics with 3D-FASTR

The increased speed of 3D-FASTR enables faster volume acquisition and presents the possibility of capturing dynamic processes volumetrically using point-scan systems. As a proof of concept, we demonstrate the utility of dynamic volumetric imaging on 4 μm beads diffusing in a 50% by weight glycerol solution. For comparison purposes, a single volume of the moving particle was acquired in 16 frame-times using the conventional stage stack approach. From the same sample, several 3D-FASTR volumes were acquired sequentially and assembled into volumes in intervals of 4 frame-times (25% $T_{vol}$) using a pattern rate of $R$ = 35 + 7/16 cy/fr.

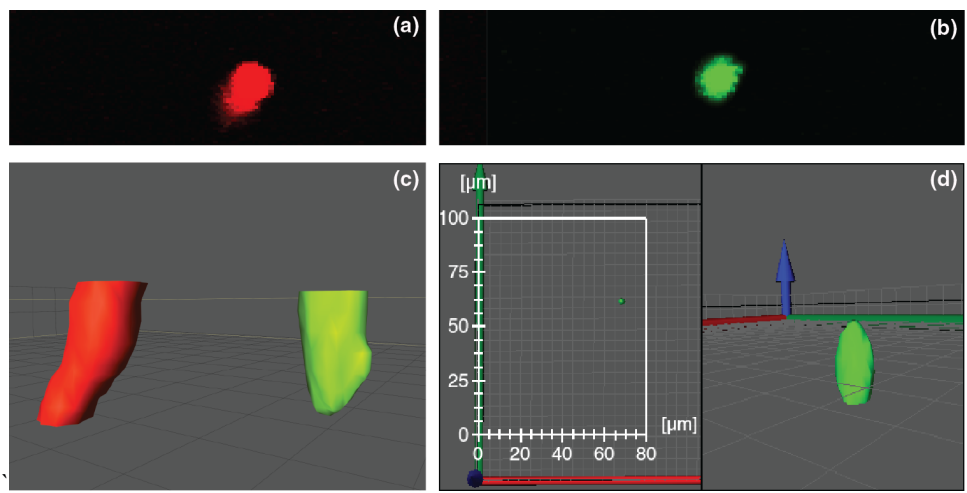

Fig. 9. Multidimensional dynamics of a 4μm fluorescent microsphere diffusing through an aqueous solution of 50% glycerol. (a). Maximum intensity projection (MIP) of bead diffusing captured using a conventional stage stack. Motion of bead during acquisition leads to a diagonal, smeared-out appearance. (b) MIP of diffusive bead captured using 3D-FASTR at 25% $T_{vol}$. Because the bead is captured with greater XZ sampling rate and in less time, there is no motion smearing or geometric distortion. (c) Two separately acquired volumes, co-rendered within the image space. The green volume was acquired at 4x speed using 3D-FASTR. The red volume was acquired at 1x speed using a conventional stage-step. The stage stack volume has a visibly tilted and kinked appearance, compared to the mostly-spherical 3D-FASTR volume. (d) Representative frame from **Visualization 1** shows diffusive motion of microsphere over time in 3D as captured by 3D-FASTR. Left panel shows XY motion of bead with scalebars. Right panel shows close 3D view during diffusion.

Fig. 9(a). shows the maximum intensity projection (MIP) of the stage stack, color-coded in red. Because the particle is moving during acquisition, it appears to change position at each image plane within the stack, leading to the appearance of diagonal smearing in the MIP image. The extent of this smearing will depend on particle speed. In contrast, the MIP shown in Fig. 9(b) demonstrates that when acquired using 3D-FASTR (color-coded in green) at 4x speed ($T_{vol}/4$), the image of the particle remains mostly circular in shape.

This result is alternatively visualized by locating the two separate volumes in the image field and rendering them together to compare their respective geometries in Fig. 9(c). As suggested by the MIP, when rendered as 3D isosurfaces, the structure of the conventional volume does not appear spherical but tilted with respect to the optical axis and visibly kinked. The 3D-FASTR volume, in contrast, retains its mostly spherical shape. Visualization 1 shows that the 3D-FASTR system captures the bead as it moves. A sample frame is shown as Fig.

9(d). This visualization shows movement of the bead from two different angles, demonstrating that the speed increases made possible by 3D-FASTR can be utilized to visualize dynamic processes that are too fast for conventional stage-stack methods which would otherwise create geometric distortions or motion-induced smearing.

## 5. Conclusion

In this work we have demonstrated a general method for improving the multi-dimensional imaging efficiency of point-scan imaging methods. The method utilizes tuning of the relative frequencies of linear scans to optimize fill efficiency to avoid oversampling and create evenly sampled multi-dimensional image spaces. This method was then demonstrated as 3D-FASTR, which added a linear axial translation to traditional 2D raster scanning LSM using an ETL. By carefully selecting the relative scan frequencies, improvement of volumetric imaging rates up to four-fold was achieved. This has a profound effect not just on the temporal resolution of these methods, but on light-dose considerations, which are always a concern in CLSM and 2P-LSM. Acquiring the same volumetric information from 25% of the pixels may lead to a four-fold increase in the viability of cells during repeated imaging. Finally, in addition to its benefits to live cell fluorescence microscopy, we believe the theory we have laid out in this work will be applicable to other multi-dimensional point scanning methods, ranging from AFM to SRS.

## Funding

National Institute of General Medical Sciences (NIGMS) of the National Institutes of Health (NIH) under award number R35GM124868.

## Disclosures

The authors declare no conflict of interest.